\documentclass{article}
\usepackage{epsfig}
\usepackage{amssymb}
\usepackage{amsmath}
\usepackage{authblk}
\usepackage{placeins}
\usepackage[utf8]{inputenc}
\usepackage[polish,english]{babel}
\usepackage[OT4]{fontenc}
\usepackage[MeX]{polski}
\usepackage{float}
\usepackage{tikz}

\title{Investigating the "kink" plot as a signal of the onset of deconfinement.}

\author{Michał Naskręt}
\providecommand{\keywords}[1]{Key words: #1}
\providecommand{\PACS}[1]{PACS numbers: #1}
\affil{University of Wroclaw, NA61/SHINE}
\date{}

\begin{document}

\maketitle

\begin{abstract}
One of the physics goals of the NA61/SHINE collaboration at the CERN Super Proton Synchrotron is to study the phase diagram of hadronic matter. To this end, a series of heavy ion collision measurements are performed. It is believed that above a certain collision energy and system size a phase transition between hadronic matter and quark-gluon plasma occurs. A number of observables has been developed to determine which of the phases was created at the early stage of the collision. This report discusses the dependence of the ratio of the mean number of produced pions to the mean number of wounded nucleons on the Fermi energy measure. For comparison with other measurements this is often presented in the form of the the "kink" plot. This plot is presented enriched with preliminary results for Ar+Sc central collisions at 13\textit{A}, 19\textit{A}, 30\textit{A}, 40\textit{A}, 75\textit{A} and 150\textit{A} GeV/c beam momentum. The results are finally compared to data from other experiments.

\end{abstract}
\keywords{NA/61 SHINE, CERN, quark-gluon plasma, onset of deconfinement}\\
\PACS{25.75.-q, 25.75.Nq}

\section{Introduction}
NA61/SHINE at the CERN SPS is a fixed target experiment which studies final hadronic states produced in interactions between various particles at different collision energies~\cite{na61det}. The search for the onset of deconfinement of strongly interacting matter is part of the experiment's physics program. Within the framework of this program measurements of p+p, Be+Be, p+Pb, Ar+Sc and Pb+Pb collisions were performed. In the nearest future measurements of Xe+La reactions are planned. The measurements will greatly benefit the two-dimensional system size and collision energy scan, which will be helpful in studying the phase transition between hadronic and deconfined phases.

\begin{figure}[H]
  \centering
    \includegraphics[width=\textwidth]{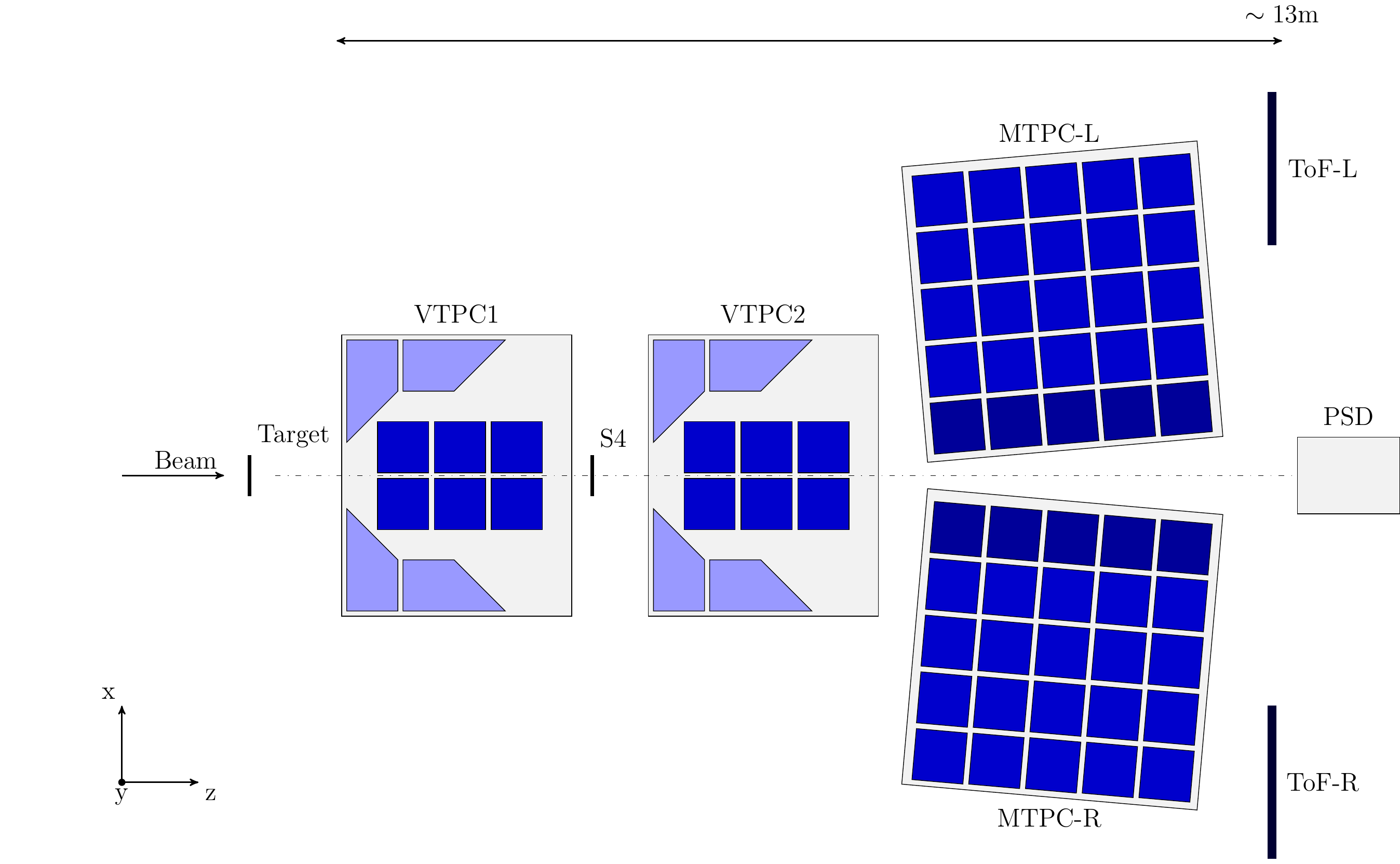}
  \caption{Experimental setup of the NA61/SHINE}
  \label{fig:setup}
\end{figure}

The NA61/SHINE detector (see Fig.~\ref{fig:setup}) is a large-acceptance hadron spectrometer located on the SPS ring at CERN. Upstream of the spectrometer are placed scintillators, Cherenkov detectors and Beam Position Detectors. They provide timing references and position measurements for the incoming beam particles. About 4 meters after the target, the trigger scintillator counter S4 detects if collisions are in the target area by the absence of a beam particle signal. The main detectors of NA61/SHINE are four large volume Time Projection Chambers used for determination of trajectories and energy loss $dE/dx$ of produced charged particles. The first two -- VTPC-1 and VTPC-2 are placed in a magnetic field to measure particles' charge and momentum. Two large TPCs (MTPC-L and MTPC-R) are placed downstream of the magnets. Time-of-flight measurements are performed by two ToF walls (ToF-L, ToF-R). The last part of the setup is the Projectile Spectator Detector (PSD) -- a zero-degree calorimeter which measures the energy of projectile spectators.

\section{The "kink" plot}
According to the Standard Model, quarks and gluons are elementary constituents of hadronic matter. Interactions between them are described by quantum chromodynamics~\cite{gellmann,zweig}. One of the most important properties of quarks and gluons is that in matter under normal conditions they are always enclosed within hadrons. This is called quark confinement. In order to transform matter from the hadronic into the deconfined phase one needs to create high temperature and system density conditions. This can be achieved experimentally in ultra-relativistic heavy ion collisions. A system that achieves the necessary conditions during the early stage of the collision forms the quark-gluon plasma~\cite{qgp:1}, which rapidly expands and freezes-out. In this later stage of the collision, quarks and gluons recombine and form a new set of hadrons.

\begin{figure}[H]
  \centering
    \includegraphics[width=0.9\textwidth]{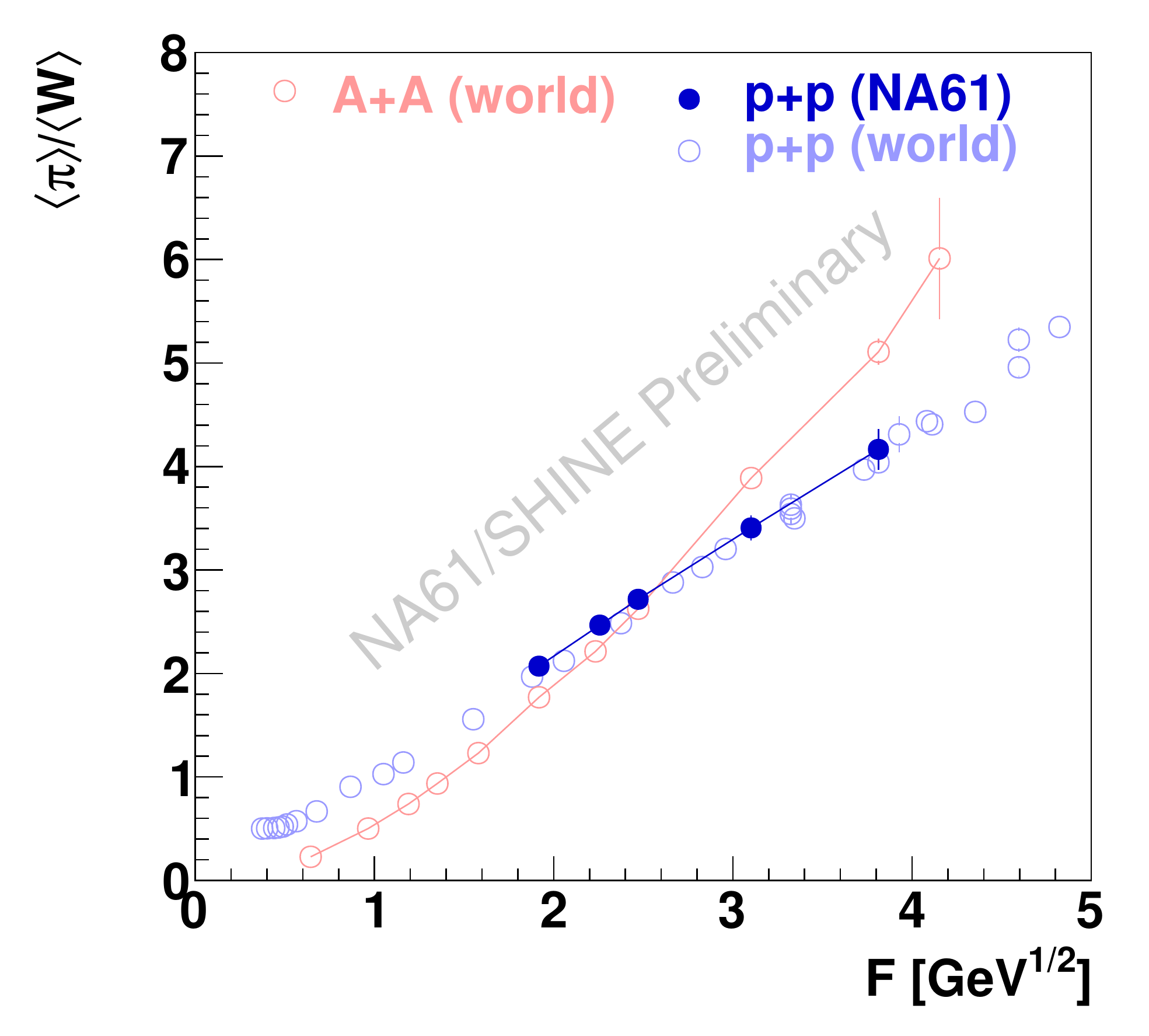}
  \caption{The "kink" plot for p-p and nucleus-nucleus collisions.}
  \label{fig:kinkRaw}
\end{figure}

Since the number of degrees of freedom is higher for the quark-gluon plasma than for confined matter, it is expected that the entropy density of the system at given temperature and density should also be higher in the first case. The majority of particles ($\sim90\%$) produced in heavy ion collisions are $\pi$ mesons. Therefore, the entropy and information regarding the state of matter formed in the early stage of a collision should be reflected in the number of produced pions normalised to the volume of the system. This intuitive argument was quantified within the Statistical Model of the Early Stage (SMES)~\cite{pedestrians}. The change of the mean number of produced pions $\langle \pi \rangle$ is often normalized to the number of wounded nucleons $\langle W \rangle$ and plotted against the Fermi energy measure $\left( F=\left[(\sqrt{s_{\text{NN}}}-2m_{\text{N}})^3/\sqrt{s_{\text{NN}}}\right]^{1/4} \right)$. As the plot resembles a linear increase with a kink of slope in the case of nucleus-nucleus collisions, it is often referred to as the "kink" plot. The plot is presented in Fig.~\ref{fig:kinkRaw}.

\section{The "kink" plot enriched with preliminary results from Ar+Sc collisions}
In order to add a new point to the "kink" plot it is necessary to calculate the mean number of pions produced in a collision and the mean number of wounded nucleons. The former is extracted from experimental data. In this paper, preliminary results obtained from Ar+Sc central collisions at 13\textit{A}, 19\textit{A}, 30\textit{A}, 40\textit{A}, 75\textit{A} and 150\textit{A} GeV/c, taken during 2015 NA61/SHINE physics data run are discussed. The number of wounded nucleons is not measured experimentally and has to be calculated using Monte Carlo simulations.

\subsection{Calculating the mean number of produced pions}
\begin{figure}[H]
  \centering
    \includegraphics[width=0.9\textwidth]{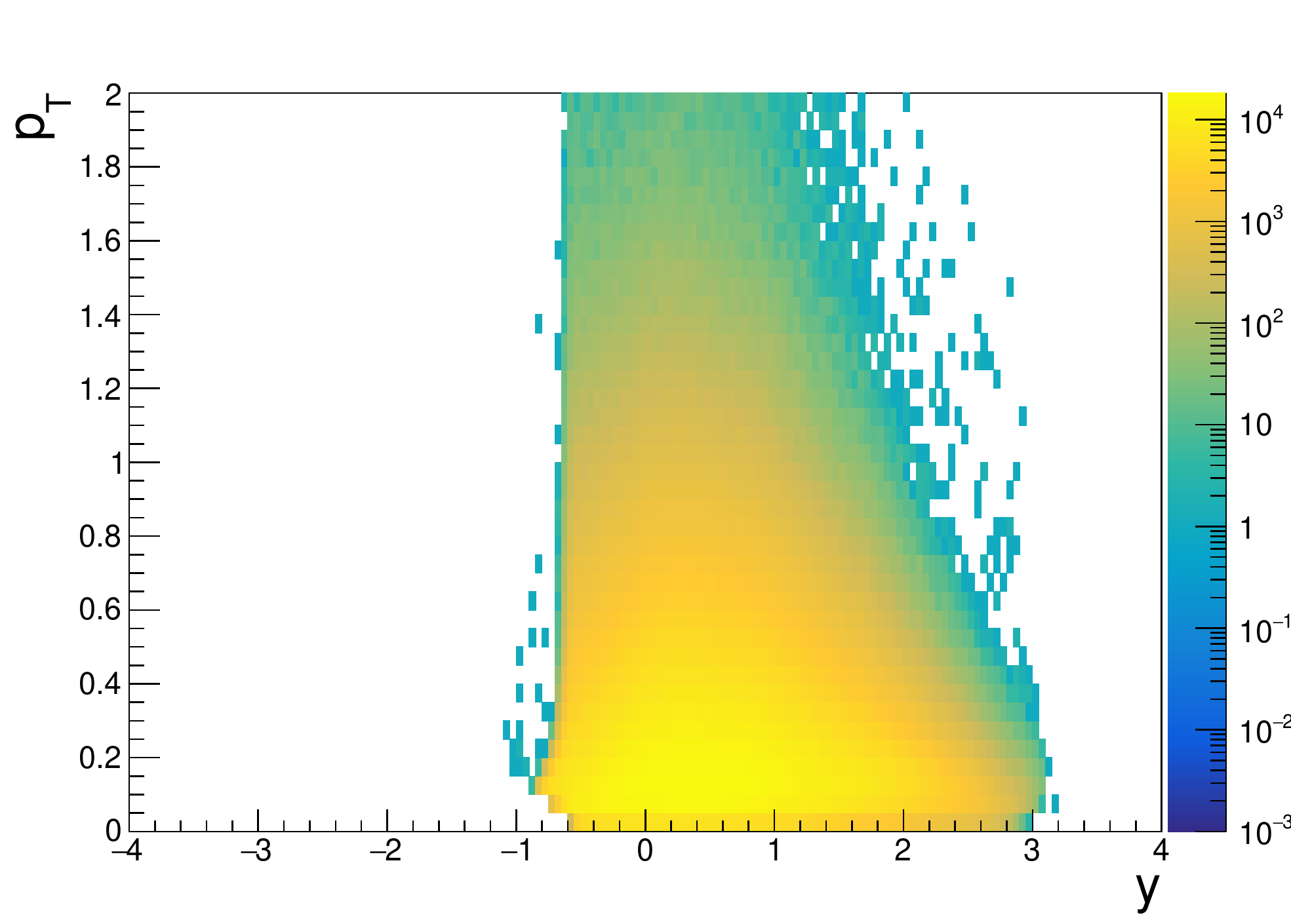}
  \caption{Example of a double differential spectrum $\frac{dn}{dydp_{\text{T}}}$.}
  \label{fig:spectrum}
\end{figure}

The starting point of the analysis described herein are double differential spectra $\frac{dn}{dydp_{\text{T}}}$ of negatively charged hadrons (see expample Fig.~\ref{fig:spectrum}).

They were obtained from reconstructed tracks applying a series of quality cuts. In order to correct for trigger and reconstruction inefficiencies, one needs to apply a Monte Carlo correction. To this end, the EPOS MC~\cite{EPOS} is used in NA61/SHINE. A large statistics of particle collisions is generated and particles are accumulated in bins $n_{\text{gen}}^{i,j}$ in transverse momentum $p_{\text{T}}$ versus rapidity. The generated data undergoes the regular reconstruction procedure and negatively charged pion selection resuling in the distribution $n_{\text{sel}}^{i,j}$. The correction factor $c^{i,j}$ is then calculated as the ratio of the two Monte-Carlo generated spectra $c^{i,j}=n_{\text{gen}}^{i,j}/n_{\text{sel}}^{i,j}$. The final experimental spectra are obtained in the following way

$$n^{i,j}=n^{i,j}_{\text{data}}c^{i,j}$$

The NA61/SHINE experimental apparatus is characterized by large, but limited acceptance. In order to estimate the mean $\pi^-$ multiplicity in the full acceptance, one needs to extrapolate the experimental data to unmeasured regions. The procedure consists of the following steps:

\begin{enumerate}
 \item \label{extrapolation:1} For fixed $y$ extrapolate the $p_{\text{T}}$ spectrum from the edge of acceptance to $p_{\text{T}}=2 GeV/c$, using the exponential form $$f(p_{\text{T}})=c p_{\text{T}} \exp\left(\frac{-\sqrt{p_{\text{T}}^2+m^2}}{T}\right)$$
 To obtain $\frac{dn}{dy}$, the measured $p_{\text{T}}$ data bins are summed and the integral of the extrapolated curve is added
 $$\frac{dn}{dy}=\sum_0^{p_{\text{T}}^{\text{max}}}dp_{\text{T}}\left(\frac{dn}{dydp_{\text{T}}}\right)_{\text{measured}}+\int_{p_{\text{T}}^{\text{max}}}^2f(p_{\text{T}})dp_{\text{T}}$$
 \item \label{extrapolation:2} The corrected rapidity spectrum is extrapolated to missing rapidity acceptance, using a sum of two symmetrically displaced Gaussians.
 $$g(y)=\frac{A_0A_{rel}}{\sigma\sqrt{2\pi}}\exp\left(-\frac{(y-y_0)^2}{2\sigma^2}\right)+\frac{A_0}{\sigma\sqrt{2\pi}}\exp\left(-\frac{(y+y_0)^2}{2\sigma^2}\right)$$
\end{enumerate}

\begin{figure}[H]
 \begin{minipage}{0.94\textwidth}
  \centering
   $\xrightarrow{\text{Extrapolation in }p_{\text{T}}}$\\
     \includegraphics[width=0.49\textwidth]{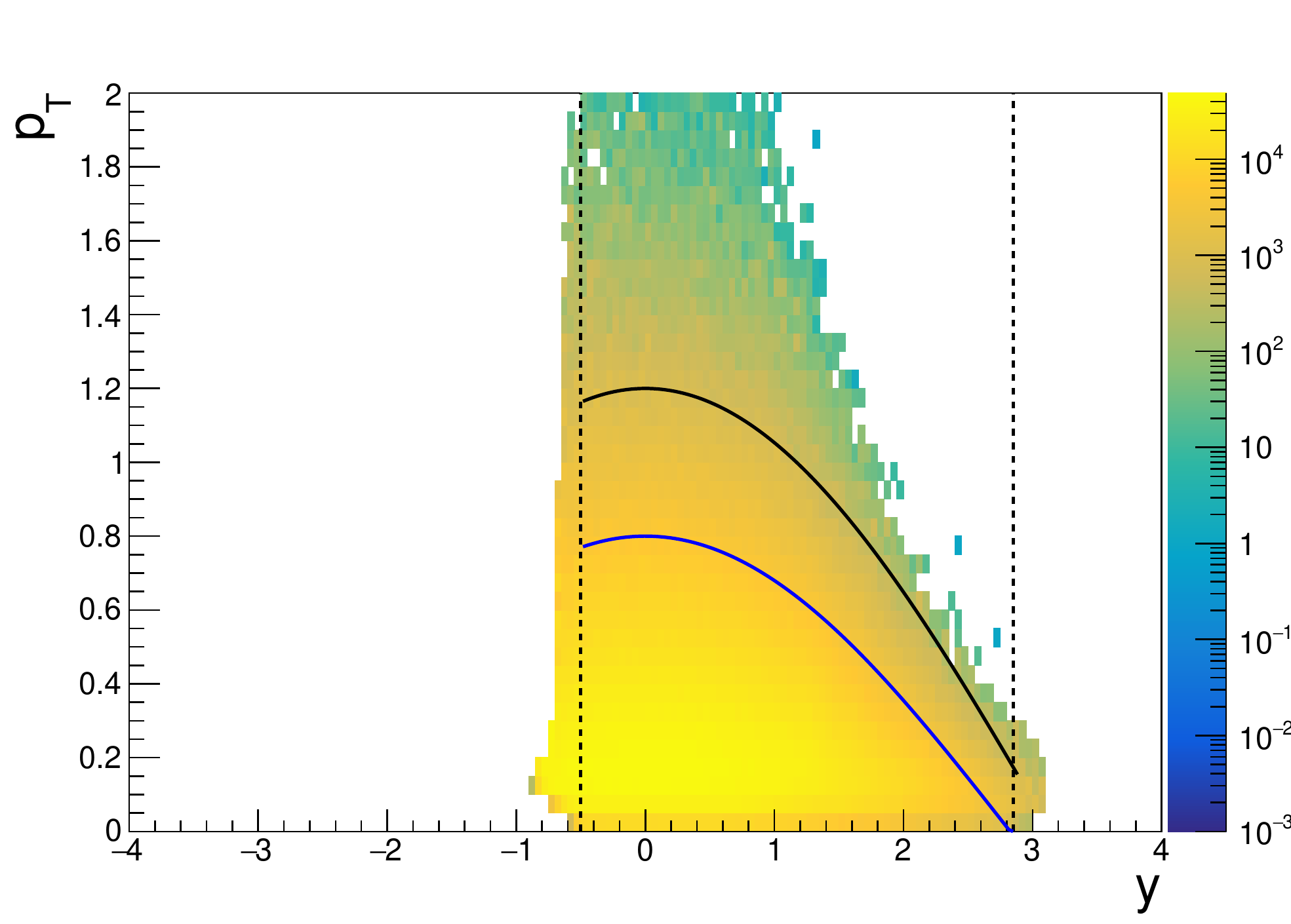}
     \includegraphics[width=0.49\textwidth]{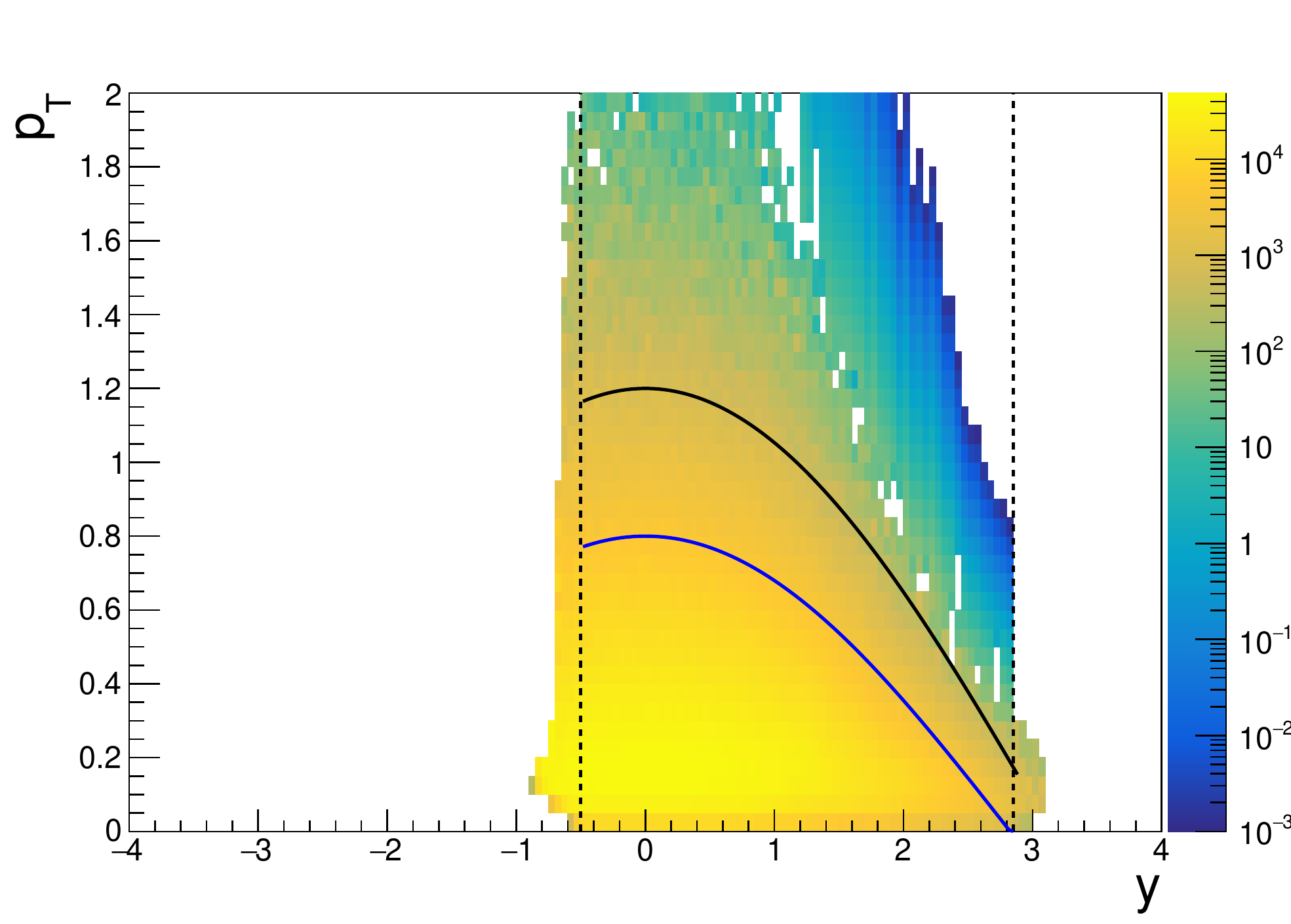}\\
     \includegraphics[width=0.49\textwidth]{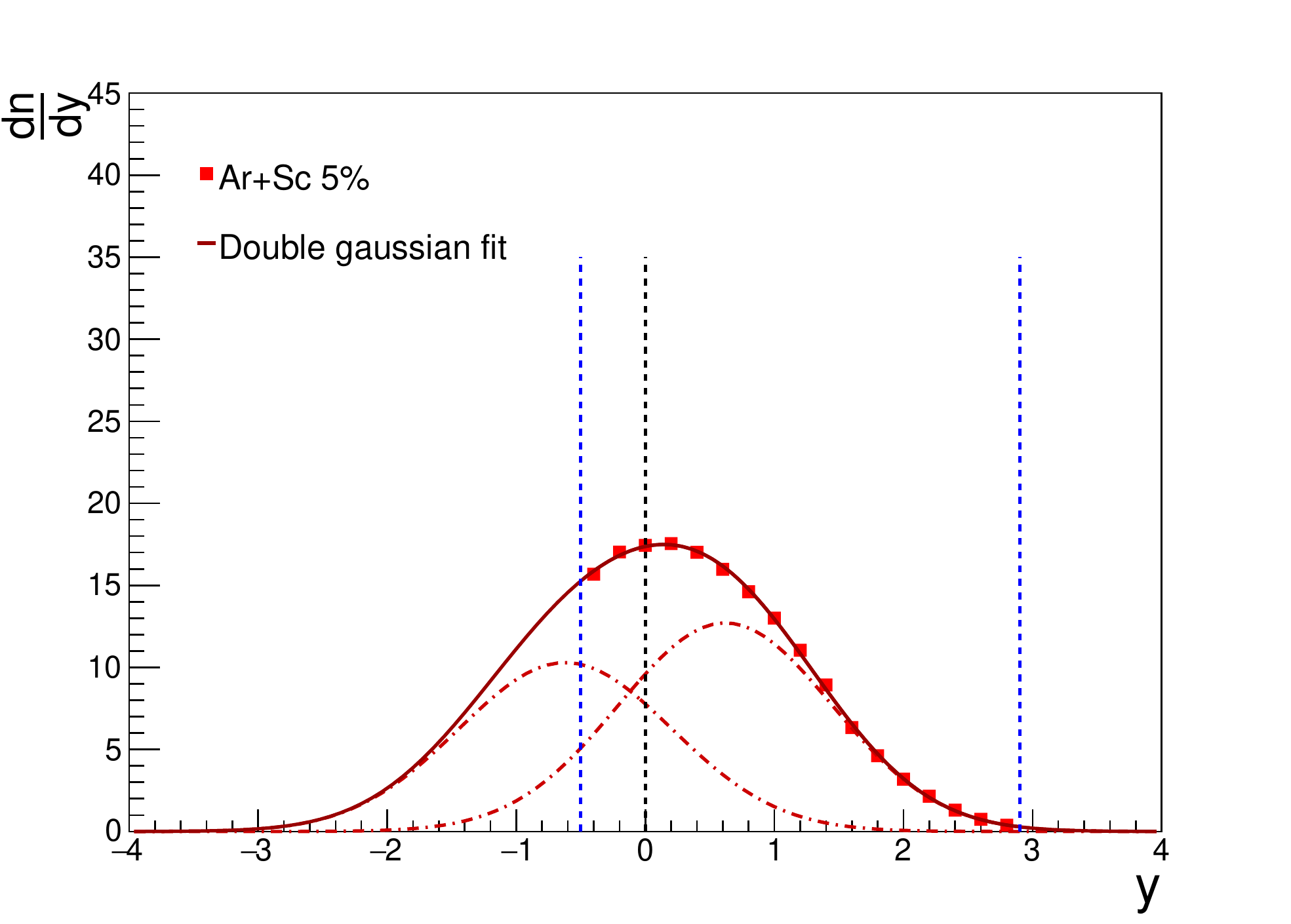}
     \includegraphics[width=0.49\textwidth]{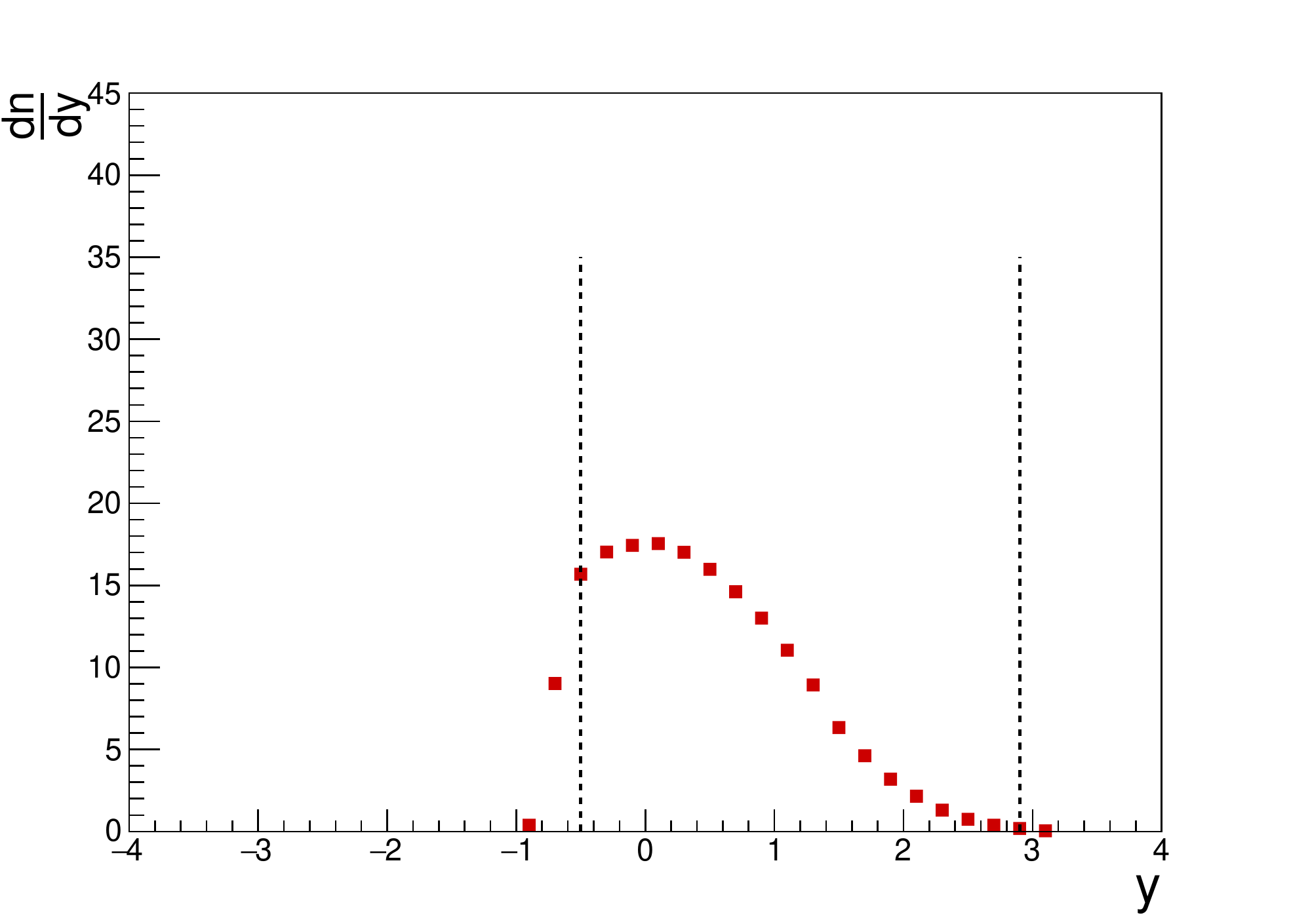}\\
   $\xleftarrow{\text{Fitting Gaussians}}$
  \end{minipage}
 \begin{minipage}{0.05\textwidth}
  $\Bigg\downarrow{\sum}$
 \end{minipage}
 \caption{Scheme of the extrapolation procedure}
 \label{fig:extrapolation}
\end{figure}

The procedure is presented schematically in Fig. \ref{fig:extrapolation}

The total mean $\pi^-$ multiplicity is given by the formula:
$$\langle \pi^- \rangle = \int_{-4}^{y_{\text{min}}}g(y)dy + \sum_{y_{\text{min}}}^{y_{\text{max}}} dy\left(\frac{dn}{dy}\right)_{\text{extrapolated in} p_{\text{T}}}+\int_{y_{\text{max}}}^4 g(y)dy$$
The results of this procedure are presented in Table~\ref{tab:piMultiplicity}. Statistical uncertainties $\sigma_{\text{stat}}(\langle \pi^{-} \rangle)$ were obtained by propagating the statistical uncertainties of $\frac{dn}{dydp_{\text{T}}}$ spectra. Systematic uncertainties $\sigma_{\text{sys}}(\langle \pi^{-} \rangle)$ are assumed to be $5\%$ based on the previous NA61 analysis of p+p collisions.

\begin{table}
 \centering
 \footnotesize
 \begin{tabular}{l|cccccc}
  Momentum [\textit{A} GeV/c] & 13 & 19 & 30 & 40 & 75 & 150\\
  \hline
  $\langle\pi^-\rangle$ & $38.46$ & $48.03$ & $59.72$ & $66.28$ & $86.12$ & $108.92$\\
  $\sigma_{\text{stat}}(\langle \pi^{-} \rangle)$ & $\pm 0.021$ & $\pm 0.021$ & $\pm 0.024$ & $\pm 0.018$ & $\pm 0.0079$ & $\pm 0.0088$\\
  $\sigma_{\text{sys}}(\langle \pi^{-} \rangle)$& $\pm 1.92$ & $\pm 2.40$ & $\pm 2.98$ & $\pm 3.31$ & $\pm 4.30$ & $\pm 5.44$
 \end{tabular}
 \caption{Mean $\pi^-$ multiplicities in the 5 \% most central Ar+Sc collisions with systematic and statistical uncertainties.}
 \label{tab:piMultiplicity}
\end{table}

\subsection{Calculating the mean number of wounded nucleons}
\begin{figure}[H]
  \centering
    \includegraphics[width=0.9\textwidth]{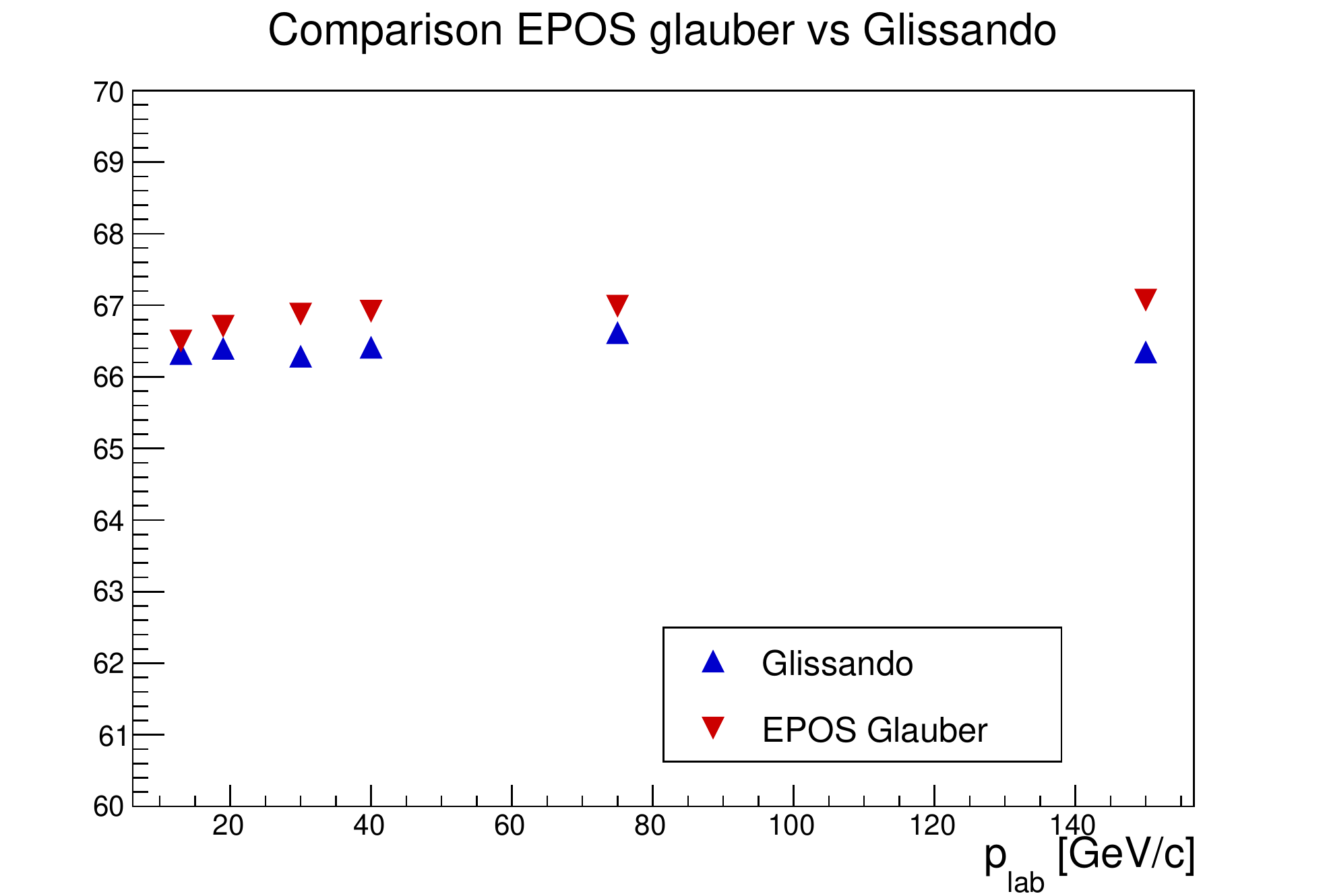}
  \caption{Comparison of Glissando and EPOS "a la Glauber" values of $\langle W \rangle$.}
  \label{fig:comparison}
\end{figure}

The number of wounded nucleons can not be measured experimentally in NA61/SHINE. It has to be calculated using Monte Carlo models. Two models were used to perform calculations -- Glissando 2.73~\cite{glauber} based on the Glauber model and EPOS 1.99 (version CRMC 1.5.3)~\cite{EPOS} using a parton ladder model. Uncertainties of $\langle W\rangle$ were not calculated and are not presented herein. A procedure to obtain a reliable number from the models has been developed. Glissando provides a value that is consistent with previous measurements and applicable to the wounded nucleon model~\cite{wounded}. EPOS, on the other hand, allows for more detailed centrality analysis and event selection. It is possible to reproduce Glauber-based values in EPOS and they are in good agreement with Glissando as shown in Fig.~\ref{fig:comparison}.

For now, the EPOS "a la Glauber" value is used with centrality selected based on the number of projectile spectators. Results are presented in the Table~\ref{tab:w}. More detailed analysis of centrality is planned as a future step of the analysis.

\begin{table}
 \centering
 \footnotesize
 \begin{tabular}{l|cccccc}
  Momentum [\textit{A} GeV/c] & 13 & 19 & 30 & 40 & 75 & 150\\
  \hline
  $\langle W\rangle$ & $66.3262$ & $66.3996$ & $66.2887$ & $66.4137$ & $66.6193$ & $66.3485$
 \end{tabular}
 \caption{$\langle W \rangle$  in the 5 \% most central Ar+Sc collisions calculated by EPOS.}
 \label{tab:w}
\end{table}

\section{Conclusions}

\begin{figure}[H]
  \centering
    \includegraphics[width=0.9\textwidth]{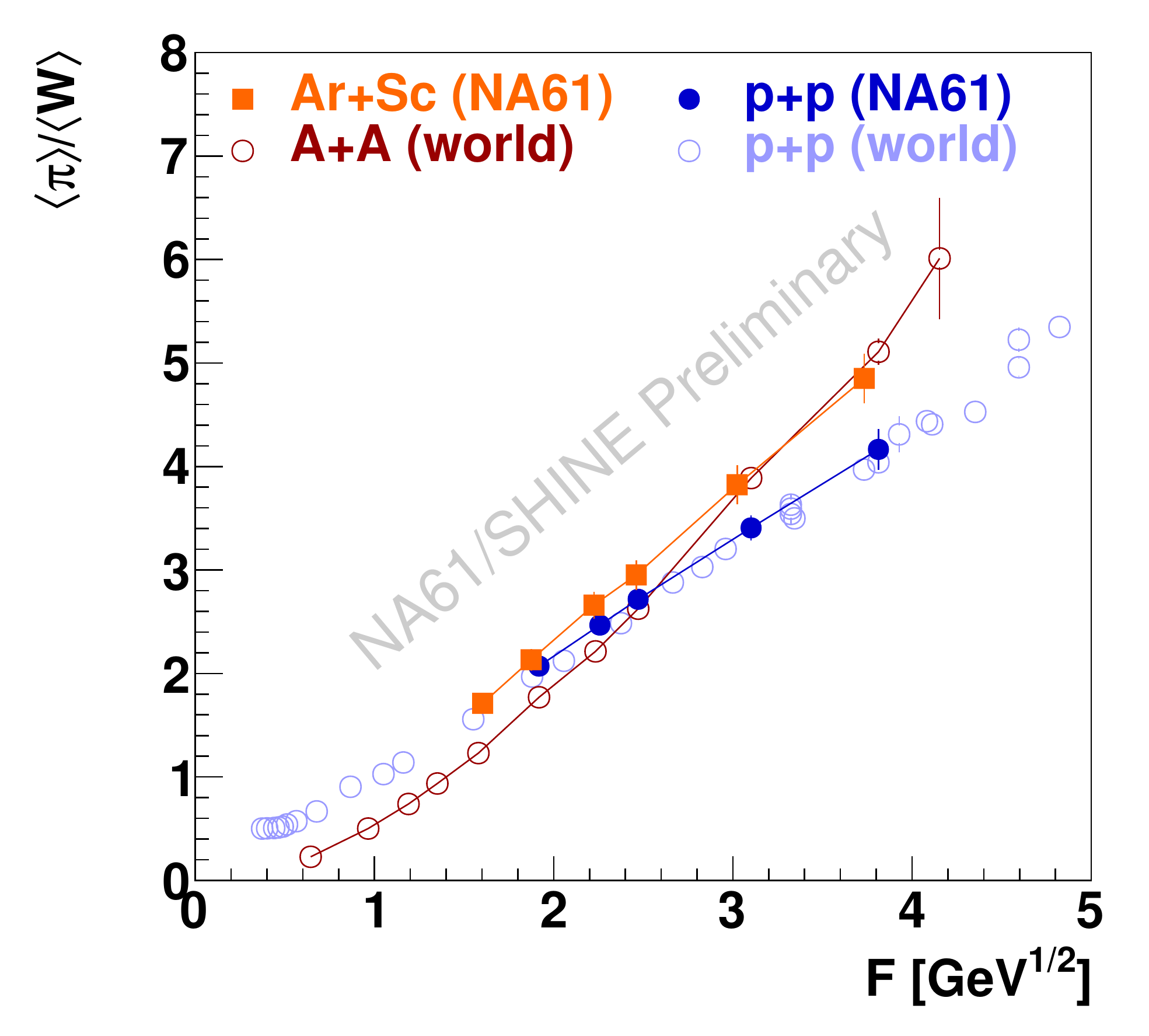}
  \caption{The "kink" plot with new measurements.}
  \label{fig:kinkNew}
\end{figure}

As for NA61/SHINE there are only results for $\langle\pi^-\rangle$ in Ar+Sc collisions, the pion multiplicity was approximated by $\langle\pi\rangle_{\text{Ar+Sc}}=3\langle\pi^-\rangle_{\text{Ar+Sc}}$. This allows to produce the preliminary version of the "kink" plot shown in Fig.~\ref{fig:kinkNew}.

From the preliminary version of the "kink" plot one can conclude that for high SPS energies Ar+Sc follows the Pb+Pb trend. On the other hand, for low  SPS energies Ar+Sc follows the p+p tendency. The plot also confirms the low-energy enhancement of $\frac{\langle \pi \rangle}{\langle W \rangle}$ measured in p+p compared to A+A collisions.

\section{Acknowledgments}

This work was partially supported by the National Science Centre, Poland grant \textit{Harmonia 7 2015/18/M/ST2/00125}.

\FloatBarrier
\bibliographystyle{unsrt}
\bibliography{bibl.bib}
\end{document}